\documentclass[]{piparticle-final}
\usepackage{graphicx}
\usepackage{amsmath}

\usepackage{cite} % To compress citation ranges
\usepackage[normalem]{ulem}  % To intoroduce strike out by Editor   \sout{text}
\usepackage{epstopdf}

\begin{document}

\volume{6}               % To be inserted by Editor
\articlenumber{060002}   % To be inserted by Editor
\journalyear{2014}       % To be inserted by Editor
\editor{E. Mizraji}   % To be inserted by Editor
\reviewers{J. Brum, Instituto de F\'isica, Facultad de Ciencias, \\ \mbox{} \hspace{3.2cm} Universidad de la Rep\'ublica, Montevideo, Uruguay.}  % To be inserted by Editor
\received{29 December 2013}     % To be inserted by Editor
\accepted{27 May 2014}   % To be inserted by Editor
\runningauthor{E. V. Bonzi \itshape{et al.}}  % To be inserted by Editor
\doi{060002}         % To be inserted by Editor

\title{Study of the characteristic parameters of the normal voices  of Argentinian speakers}
\author{E. V. Bonzi,\cite{inst1,inst2}\thanks{E-mail: bonzie@famaf.unc.edu.ar}\hspace{0.5em}  
        G. B. Grad,\cite{inst1}\thanks{E-mail: grad@famaf.unc.edu.ar}\hspace{0.5em}   
        A. M. Maggi,\cite{inst3}\thanks{E-mail: alicia.maggi@hotmail.com}\hspace{0.5em}  
	M. R. Mu\~n\'oz\cite{inst3}\thanks{E-mail: eudaimonia13@hotmail.com}}

\pipabstract{
The voice laboratory permits to study the human voices using a method that is 
objective and  noninvasive.  
In this work, we have studied the parameters of the human voice  such as pitch, formant, jitter, shimmer and harmonic-noise ratio 
 of a group of young people.  This statistical information of parameters is obtained from Argentinian speakers.

%, and have not been studied this way before. 
}

\maketitle
\blfootnote{
\begin{theaffiliation}{99}
   \institution{inst1} Facultad de Matem\'atica, Astronom\'{\i}a y F\'isica, Universidad Nacional de C\'ordoba, Ciudad Universitaria, 5000 C\'ordoba, Argentina.
   \institution{inst2} Instituto de F\'{\i}sica Enrique Gaviola (CONICET), Ciudad Universitaria, 5000 C\'ordoba, Argentina.
   \institution{inst3} Escuela de Fonoaudiolog\'{\i}a, Facultad de Ciencias M\'edicas, Universidad Nacional de C\'ordoba, Ciudad Universitaria, 5000 C\'ordoba, Argentina.
\end{theaffiliation}
}

\section{Introduction}

The voice is a multidimensional  phenomenon that must be evaluated using special tools for determining acoustic parameters.
These parameters are: the pitch or voice tone, the timbre, considered as the personality of the 
voice that is particular of each person (determined by fundamental frequency, its harmonics and formants) and the degree of hoarseness.

During sustained vibration, the vocal fold will exhibit variations of fundamental frequency and amplitude; 
these phenomena are called ``frequency perturbation'' (jitter) and ``amplitude perturbation'' (shimmer).
They reflect fluctuations in tension and biochemical characteristics of the vocal folds, as well as 
variation in their neural control and the physiological properties of the individuals voices.

The acoustic analysis is one of the major advances in the study of voice, increasing the accuracy of diagnosis in this area.
Normal values as standards are important and necessary to guide voice professionals.

There are not many studies performed for the Latin languages \cite{rodri10, cerve01, munoz03}. However, there are several of them for the English language, 
such as those in Refs. \cite{houri07, white01, white99, cole71, bennett81}.

In the same way, the software used for voice therapy is in general designed for other languages than Spanish. A comparison has been made, though, between the two vowel systems of English and Spanish (the variation spoken in Madrid, Spain), which triggered relatively large versus small vowel inventories \cite{bradlow94}.
That is the reason why we consider it is very important and necessary to produce more results for the Spanish speaking population.

We analyzed 72 audio files of female and male voices from an Argentinian Spanish speaking population to obtain the acoustical parameters using 
the Praat program~\cite{Praat}.
Our data were compared to Bradlow~\cite{bradlow94}, Hualde~\cite{hualde05} and Casado Morente~\textit{et al.}~\cite{casado01}. 
The pitches measured were lower than expected and the First formant of the /a/ and /u/ vowels is higher than the published data.
Additionally, the Harmonic to Noise Ratio (HNR) values discriminated per vowel are presented.

\section{Measurement methodology}

Pitch, First and Second formants, Jitter, Shimmer and Harmonic to Noise Ratio (HNR)͒ are the cornerstones of acoustic measurement of 
voice signals, and are often regarded as  indices of the perceived quality of both normal and pathological voices~\cite{krei05}.

\begin{figure}%[th]
%\begin{center}
\includegraphics[width=0.45\textwidth]{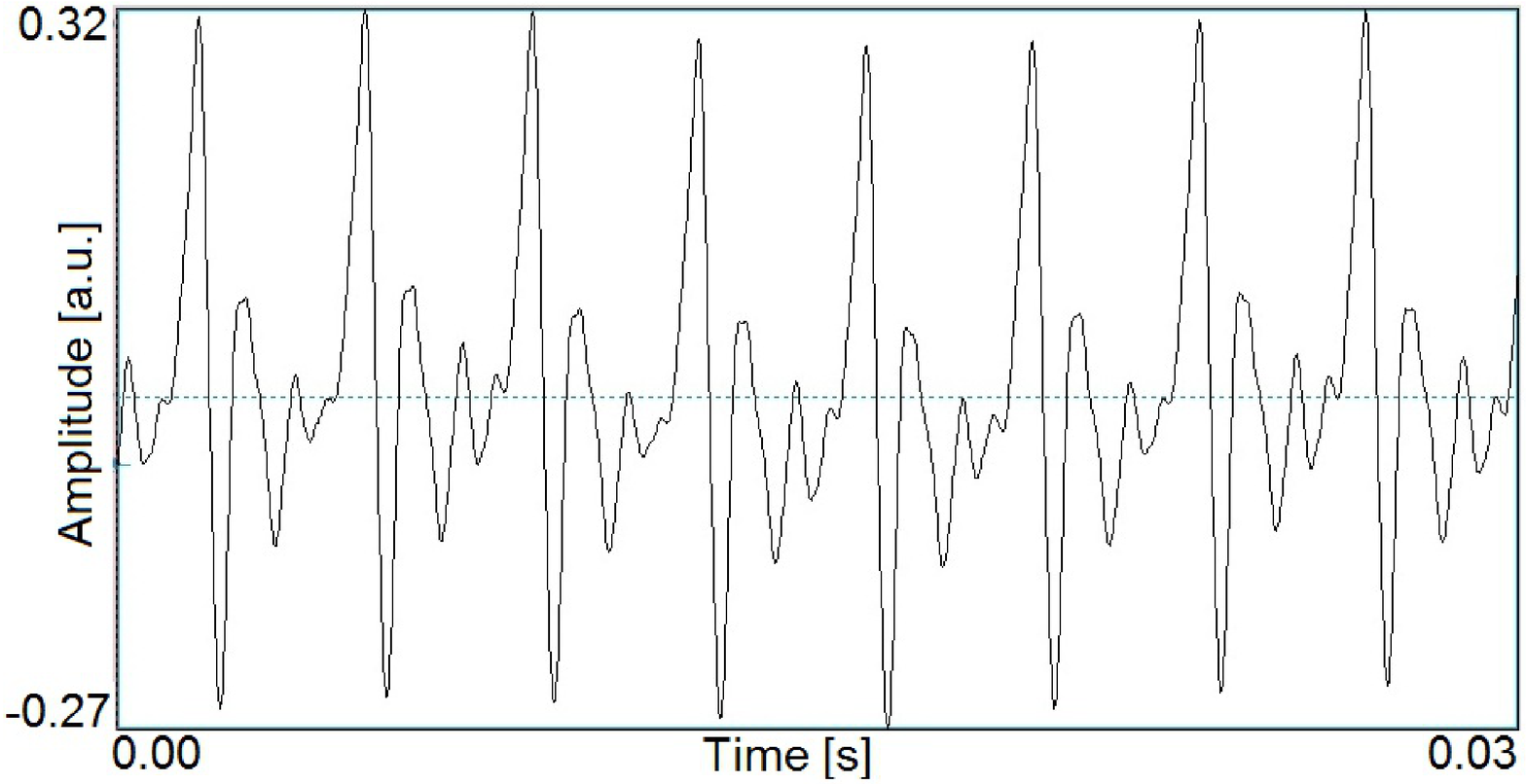}
%\end{center}
\caption{Wave shape of the /a/ sound.} \label{Oa}
\end{figure}

\begin{figure}%[th]
%\begin{center}
\includegraphics[width=0.45\textwidth]{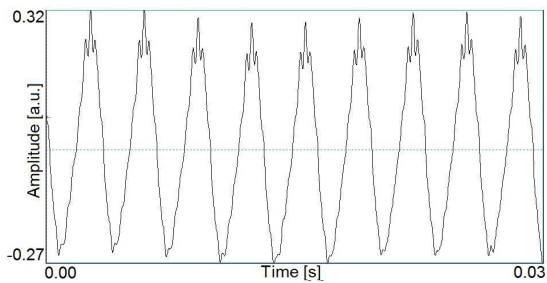}
%\end{center}
\caption{Wave shape of the /i/ sound.} \label{Oi}
\end{figure}

In this work, we  analyzed the audio files from the five Spanish vowels produced by 72 female and male individuals, 
in order to study the parameters previously mentioned.
The individuals are Argentinian university students  whose ages range between 20 and 30, coming from different regions without any special geographical distribution.

The voices were recorded using a Behringer C-1U (USB) cardioid microphone and a notebook. 

The microphone was placed at a distance of 10 cm respect to the mouth of the subjects while they
 were pronouncing the vowels with an intensity and tone that was comfortable in an  acoustically treated room. Each sound was
sustained for, at least, five seconds. 

The Praat program, commonly used in linguistics for the scientific analysis 
of the human voice~\cite{Praat}, was used to record, analyze the wav files and
obtain all the parameters presented in this work. A sample rate of 44100 Hz was used to record the sound file.

\begin{figure}
%\begin{center}
\includegraphics[width=0.45\textwidth]{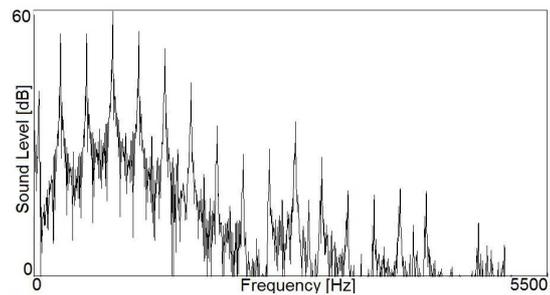}
%\end{center}
\caption{Harmonics of the /a/ vowel.} \label{Fa}
\end{figure}

\begin{figure}
%\begin{center}
\includegraphics[width=0.45\textwidth]{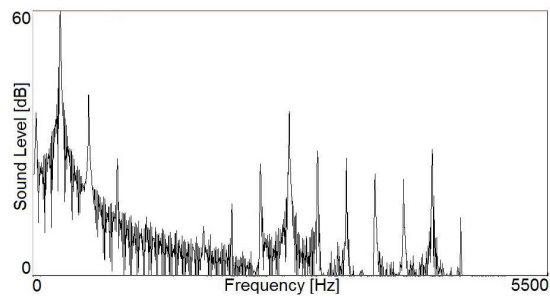}
%\end{center}
\caption{Harmonics of the /i/ vowel.} \label{Fi}
\end{figure}

The wave  shapes of the  sounds corresponding to /a/ and /i/ vowels are shown in Figs. \ref{Oa} and \ref{Oi}.
In Figs. \ref{Fa} and \ref{Fi}, the harmonic components obtained by applying Fourier 
Transform to the respective vowel signal are shown.

\textbf{Pitch}

The pitch is a perceptual attribute of sound closely related to frequency, being this perception a subjective notion.

In  psychoacoustics, the pitch is related to the fundamental frequency of vibration of the vocal cords,
allowing the perception of the tone frequency.
 
Nevertheless, for Praat program \cite{Praat}, the pitch is coincident with the fundamental harmonic of the wave and we used this definition in this work.

This parameter depends on gender, being higher for women and lower for men.

\pagebreak

\textbf{Formants}

The voice is created in the vocal cord, shaped as complex sound with harmonics and modified in the vocal tract by the resonating frequencies. Then, the amplitude of harmonics frequencies are enveloped forming a spectrum of energy, the peaks or maximum observed in these spectra are named ``formants.''
Consequently, a formant is a concentration of acoustic energy around a particular frequency in the speech wave.
There are several formants, each one at a different frequency corresponding to a resonance in the vocal tract,  and especially the first two are related to the movement of the tongue.  
The high-low magnitude of the First one (F1) is inversely related to the up-down tongue position 
and the Second formant (F2) is related to the front tongue position.

\textbf{Jitter and Shimmer}

The naturalness factor of sustained vowels is attributed to a fundamental frequency and the signal amplitude.
 Still there are unwanted variations in time of the sound signal properties in the voice production.

While jitter indicates the variability or perturbation of fundamental frequency, shimmer refers to the same perturbation but, in this case, related to amplitude of sound wave, or intensity of vocal emission. 
Jitter is affected mainly  by lack of control of vocal fold vibration and shimmer  by reduction of glottic resistance
 and mass lesions in the vocal folds, which are related  to the presence of noise at emission and breathiness ~\cite{Praat, wertzner05}.

\textbf{Harmonic to Noise Ratio - HNR}

The amount of energy conveyed in the fundamental frequency ($f_0$) and its harmonics, divided by the energy in noise frequencies, is defined as the harmonic-to-noise ratio. Frequencies that are not integer multiples of $f_0$ are 
regarded as noise. This parameter is related to the perception of vocal roughness and hoarseness ~\cite{Praat}. 

Normal voices have a low level of noise and high HNR. On the contrary, the degree of hoarseness increases the noise component and decreases HNR.

\section{Results and Discussion}

\begin{figure}
%\begin{center}
\includegraphics[width=0.45\textwidth]{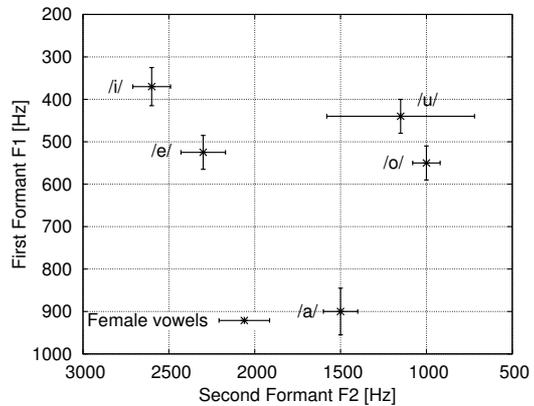}
%\end{center}
\caption{Female formant chart.} \label{CF}
\end{figure}

\begin{figure}
%\begin{center}
\includegraphics[width=0.45\textwidth]{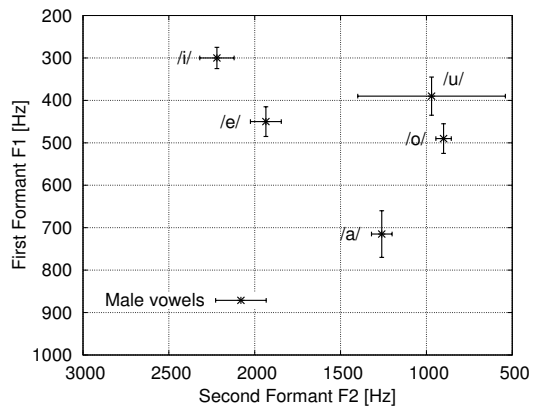}
%\end{center}
\caption{Male formant chart.} \label{CM}
\end{figure}

The measured data were processed statistically and the results are shown in the Tables~\ref{f0},~\ref{SJfemale}, 
 \ref{SJmale}, \ref{HNR} and Figs. \ref{CF} and \ref{CM}.
%average of this Spanish speaking region is shown in the tables.

The pitches for female and male individuals are shown in Table~\ref{f0}. We  used the minimum and maximum values to 
address the dispersion instead of the standard deviation because the data distribution was not normal.
Our values are in general lower for both genders compared to the published data~\cite{bradlow94, hualde05, casado01}. 

Tables~\ref{FChfemale} and \ref{FChmale} show the First and Second formants values and Figs.~\ref{CF} and \ref{CM} show the chart
 of formants corresponding to female and male populations obtained in this work. 
 
We  have compared our male results with formant data of male Spanish speakers published by Bradlow \cite{bradlow94}.

In general, the First (F1) and Second (F2) formants values are comparable to the published ones.

In particular, the F1 formants for the /a/ and /u/ vowels are higher than the reported ones, 12 and 21 \%, respectively.

The Second formant, F2, for the /o/ vowel is lower than Bradlow by 12 \%.

On the other hand, we  cannot compare our female formant values with published results because we could not find results for female individuals in the literature.
Comparing female versus male F1 formants, we  observed that most of them are higher by 20 \% but in the case of the /o/ vowel the difference is 11 \%.

Comparing F2 formants, the female values are higher than the male ones, reaching almost the 25 \% for /a/ and /i/ vowels.

Furthermore, the F2 of the /u/ vowel in our samples show an important scatter for both genders, female and male.

In the Tables~\ref{SJfemale} and \ref{SJmale}, the obtained Jitter and Shimmer values for each vowel are shown.
They are comparable to the Jitter and Shimmer averages obtained by Casado Morente~\textit{et al.}~\cite{casado01} in a study that involves
 a group of normal people.
In our work, we  have observed that the Jitter and the Shimmer values of the /a/ vowel are bigger than the corresponding ones of the other vowels.

Finally, the HNR results, see Table~\ref{HNR}, are according to the average value presented by Casado Morente~\textit{et al.}~\cite{casado01}. 
However, we could not find in the bibliography the HNR values for each of the five Spanish vowels, so we had to make the comparison  with the average of 
them. 
In the present work, we  have found that the vowels show an increasing HNR value from /a/ to /u/, meaning that  /u/ has better signal to noise
ratio than the other vowels.

 \begin{table}
\centering   
\begin{tabular}{ c  c  c }
\hline\hline
%\multicolumn{3}{|c|}{Pitch [Hz]} \\ \cline{1-3}
       &  Female & Male \\ \hline%\hline 
Maximum &  314  &  196    \\
Medium  &  225  &  128    \\
Minimum &  155  &   85    \\
\hline\hline
\end{tabular} 
\caption{Pitch values of female and male subjects in Hz.}
\label{f0}
\end{table}

\begin{table}
\centering
\begin{tabular}{ c  c  c }
\hline\hline
%Gender  & \multicolumn{2}{c|}{Female} \\ \cline{1-3}
Vowels  &  F1 [Hz] & F2 [Hz] \\ \hline%\hline
$ \it{/i/}$ & 370 $\pm$  45 & 2600 $\pm$ 110   \\
$ \it{/e/}$ & 525 $\pm$  40 & 2300 $\pm$ 130   \\
$ \it{/a/}$ & 900 $\pm$  55 & 1500 $\pm$ 100   \\
$ \it{/o/}$ & 550 $\pm$  40 & 1000 $\pm$ 80   \\
$ \it{/u/}$ & 440 $\pm$  40 & 1150 $\pm$ 430   \\ 
\hline\hline
\end{tabular}
\caption{First and Second formant of female.}
\label{FChfemale}
\end{table}

\begin{table}
\centering
\begin{tabular}{ c  c  c }
\hline\hline
%Gender  & \multicolumn{2}{c|}{Male} \\ \cline{1-3}
Vowels  &  F1 [Hz] & F2 [Hz] \\ \hline%\hline
$ \it{/i/}$ & 300 $\pm$ 25  & 2220 $\pm$ 100   \\
$ \it{/e/}$ & 450 $\pm$ 35  & 1935 $\pm$ 90   \\
$ \it{/a/}$ & 715 $\pm$ 55  & 1260 $\pm$ 60   \\
$ \it{/o/}$ & 490 $\pm$ 35  & 900  $\pm$ 45   \\
$ \it{/u/}$ & 390 $\pm$ 45  & 970  $\pm$ 430   \\ 
\hline\hline
\end{tabular}
\caption{First and Second formant of male.}
\label{FChmale}
\end{table}

\begin{table}
\centering
\begin{tabular}{ c  c  c }
\hline\hline
%Gender  & \multicolumn{2}{c|}{Female} \\ \cline{1-3}
Vowels  &  Shimmer Local [\%] & Jitter Local [\%] \\ \hline%\hline
$ \it{/a/}$ & 2.7 $\pm$ 1.1 & 0.31 $\pm$ 0.10   \\
$ \it{/e/}$ & 2.1 $\pm$ 0.7 & 0.28 $\pm$ 0.08   \\
$ \it{/i/}$ & 2.2 $\pm$ 0.6 & 0.29 $\pm$ 0.07   \\
$ \it{/o/}$ & 2.0 $\pm$ 0.7 & 0.26 $\pm$ 0.11   \\
$ \it{/u/}$ & 2.1 $\pm$ 0.7 & 0.27 $\pm$ 0.09   \\ 
\hline\hline
\end{tabular}
\caption{Shimmer and Jitter of female subjects.}
\label{SJfemale}
\end{table}

\begin{table}
\centering
\begin{tabular}{ c  c  c }
\hline\hline
%Gender  & \multicolumn{2}{c|}{Male} \\ \cline{1-3}
Vowels  &  Shimmer  Local [\%] & Jitter  Local [\%] \\ \hline%\hline
$ \it{/a/}$ & 3.0  $\pm$ 0.9  & 0.36 $\pm$ 0.10  \\
$ \it{/e/}$ & 2.3  $\pm$ 0.8  & 0.33 $\pm$ 0.09  \\
$ \it{/i/}$ & 2.3  $\pm$ 0.7  & 0.28 $\pm$ 0.08   \\
$ \it{/o/}$ & 2.2  $\pm$ 0.8  & 0.29 $\pm$ 0.10  \\
$ \it{/u/}$ & 2.3  $\pm$ 0.9  & 0.25 $\pm$ 0.07  \\ 
\hline\hline
\end{tabular} 
\caption{Shimmer and Jitter of male subjects.}
\label{SJmale}
\end{table}

\begin{table}
\centering
\begin{tabular}{ c  c  c }
\hline\hline
%\multicolumn{3}{|c|}{HNR [dB]} \\ \cline{1-3}
Vowels &  Female & Male   \\ \hline%\hline 
$\it{/a/}$ & 21 $\pm$ 3 &  20 $\pm$ 2 \\
$\it{/e/}$ & 20 $\pm$ 2 &  21 $\pm$ 2 \\
$\it{/i/} $ &  22 $\pm$ 3 &  22 $\pm$ 2 \\
$\it{/o/} $ &  25 $\pm$ 3 &  24 $\pm$ 3 \\
$\it{/u/} $ &  25 $\pm$ 4 &  25 $\pm$ 3 \\
\hline\hline
\end{tabular} 
\caption{Harmonic to Noise Ratio of female and male subjects in dB.}
\label{HNR}

\end{table}

\section{Concluding remarks}

The objective of this research was to measure acoustical properties of the Spanish voices
of Argentinian speakers.

These voice parameters are generally assessed subjectively by several authors. 
This form of perceptual analysis of voice has significant limitations and the subtle interpretative judgments of verbal classifications may not be accurate. 

The differences we found in the parameters of the vowels measured in a group of people from Argentina compared to the parameters obtained from Spanish speaking people living in Spain suggests the region of study has an important influence in the results, as expected.

This kind of studies are very useful to compare the properties of normal and pathological voices of people from different regions.

It is necessary to test the same parameters in female Spanish speakers as well.

Such work should be performed in larger quantities and should be extended to other countries or regions of Latin America, especially where different ethnic groups can be found.

\end{document}